# Triangular Snowflakes:
## Growing Structures with Three-fold Symmetry using a Hexagonal Ice Crystal Lattice

Kenneth G. Libbrecht

Department of Physics, California Institute of Technology
Pasadena, California 91125, kgl@caltech.edu

**Abstract.** Snow crystals growing from water vapor occasionally exhibit morphologies with three-fold (trigonal) symmetry, even though the ice crystal lattice has a molecular structure with six-fold symmetry. In extreme cases, thin platelike snow crystals can grow into faceted forms that resemble simple equilateral triangles. Although far less common than hexagonal forms, trigonal snow crystals have long been observed both in nature and in laboratory studies, and their origin has been an enduring scientific puzzle. In this paper I describe how platelike trigonal structures can be grown on the ends of slender ice needles in air with high reliability at -14 C. I further suggest a physical model that describes how such structures can self-assemble and develop, facilitated by an edge-sharpening instability that turns on at a specific combination of temperature and water-vapor supersaturation. The results generally support a comprehensive model of structure-dependent attachment kinetics in ice growth that has been found to explain many of the overarching behaviors seen in the Nakaya diagram of snow crystal morphologies.

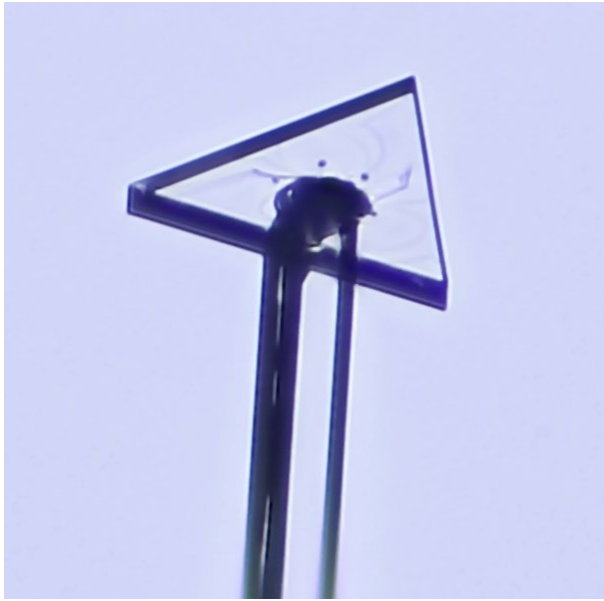 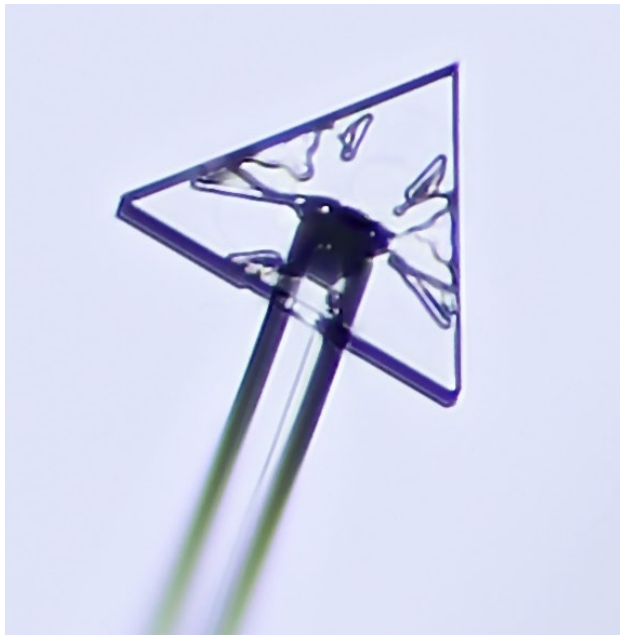

Figure 1: This paper considers the formation and development of snow crystals with large-scale trigonal symmetry, such as the two laboratory-grown examples shown above. Ignoring minor surface markings, both these crystals are essentially equilateral-triangular ice plates growing on the ends of c-axis ice columns, and each structure consists of a single crystal of hexagonal ice Ih bounded by basal and prism facets.

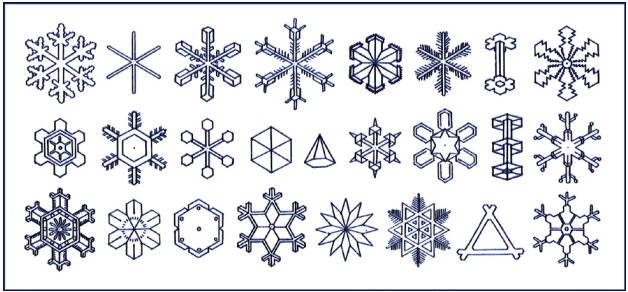

Figure 2: Snowflake sketches by William Scoresby [1820Sco] included one specimen with three-fold symmetry (next to last in the diagram), which may be the first recorded observation of a trigonal snow crystal.

## ❋ Observations of Trigonal Snow Crystals

The earliest mention of trigonal snow crystals I have found is by William Scoresby [1820Sco], who included one example in his sketches of arctic snow crystals shown in Figure 2. Wilson Bentley presented many trigonal crystals in his photographic album [1931Ben], devoting a small morphological section to platelike forms with three-fold symmetry. Nakaya presented some photographic examples as well [1954Nak], although many of these were split stars that had broken into three-branched stars [2021Lib]. While the latter do exhibit three-fold symmetry, their formation involves a breakage step that is not present in the trigonal snow crystals discussed below. (For this reason, I exclude broken split stars from the discussion of true trigonal forms below.) Murray et al. [2015Mur] have summarized many subsequent observations of trigonal snow crystal forms in the Earth's atmosphere [2002Hal, 2013Shu, 2021Mag].

Trigonal snow crystals are not especially difficult to find in natural snowfalls in temperate climates if one examines numerous small specimens, but of course they are far less common than hexagonal forms. As illustrated in Figure 3, simple plates with surface markings are the most common trigonal crystals, with the most basic design exhibiting alternating long and short prism facets. In my experience, trigonal forms in nature tend to cluster in time, meaning that some snowfalls bring enough to provide for multiple observations, while other snowfalls provide essentially none. This basic observation suggests that some atmospheric conditions are especially conducive to trigonal formation, a statement than can be made for most snow crystal morphologies.

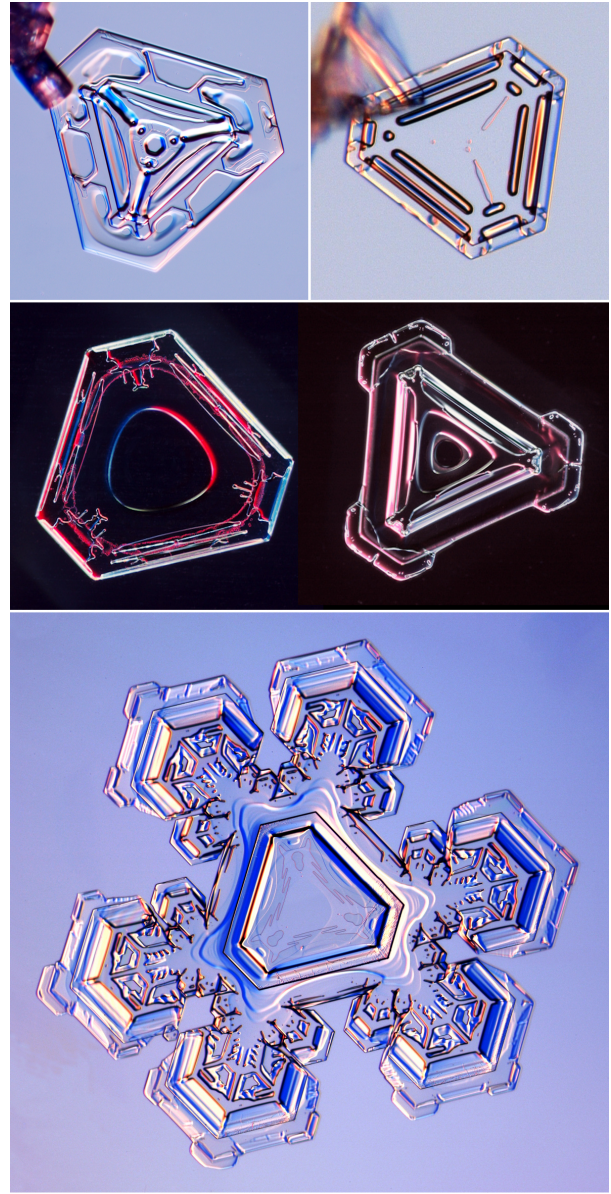

Figure 3: Photographs of several natural trigonal snow crystals taken by the author. The small platelike forms (top) range in size from 0.6 mm to 1.4 mm, while the branched crystal (bottom) measures about 2.5 mm from tip to tip.

## Nomenclature for Simple Platelike Crystals

The focus of this paper is on simple platelike trigonal forms (without branching), and for these crystals I have found it useful to define a "hexagonality" parameter

$$H_1 = L_1/L_6$$

and a "triangularity" parameter

$$T_1 = 1 - L_3/L_4$$

for each individual specimen, where the six prism facets are ordered by length from $L_1$ (shortest) to $L_6$ (longest). A perfect hexagon then has $(H_1, T_1) = (1,0)$, while a perfect equilateral triangle has $(H_1, T_1) = (0,1)$. For a typical sample of simple snow-crystal plates, the vast majority are clustered near $(H_1, T_1) \approx (1,0)$ when plotted in the $(H_1, T_1)$ plane, indicating basic hexagonal forms. As a rule of thumb, any crystal with $H_1 > 0.85$ looks essentially hexagonal by eye.

If one removes high-$H_1$ crystals from the sample, the remaining crystals tend to be clustered on a line with $H_1 + T_1 = 1$, as illustrated in Figure 4. I call this the "trigonal line," as platelike crystals near this line generally exhibit three long and three short prism facets that alternate around the perimeter. Figure 5 shows some examples of trigonal platelike crystals (the top six photos) and irregular crystals (the lower six photos). After excluding hexagonal crystals from the sample, trigonal crystals are the most common, while irregular crystals are rare.

By placing crystals on the $(H_1, T_1)$ plane, it becomes clear that the most common simple, platelike forms span a continuum along the trigonal line, from perfect hexagons at one end to perfect equilateral triangles at the other end. With this continuum in mind, I often speak of "hexagons" with $(H_1, T_1) \approx (1,0)$, "triangular" crystals with $(H_1, T_1) \approx (0,1)$,

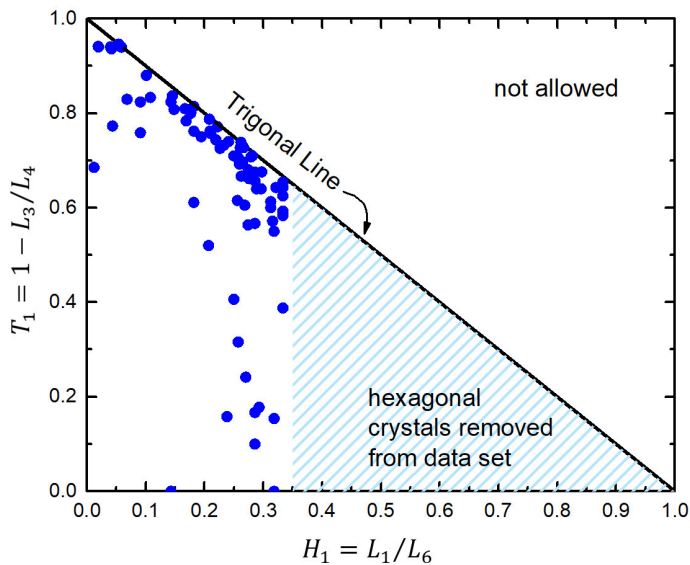

Figure 4: This graph shows a set of snow-crystal plates plotted on the $(H_1, T_1)$ plane. Each data point represents a single crystal grown at -10 C in a free-fall convection chamber [2009Lib4]. Photographic examples from this data set are shown in Figure 5. Crystals above the trigonal line are not allowed from basic geometrical considerations, and data points for hexagonal (high $H_1$) crystals were excluded from this plot. These data illustrate that most simple platelike crystals with low $H_1$ tend to cluster near the trigonal line, where the morphology exhibits a clear three-fold symmetry (see Figure 5). This clustering suggests that some deterministic physical mechanism is responsible for promoting the growth of trigonal forms [2009Lib4]. The irregular morphologies far from the trigonal line (see Figure 5) may be influenced by dislocations, heterogeneous nucleation processes from impurities, collisions between free-falling crystals, or other imperfections that may arise during growth.

and I use "trigonal" crystals to mean anywhere near the trigonal line with $H_1 \lesssim 0.85$.

A basic statistical analysis [2009Lib4] reveals that trigonal crystals are indeed much more common than one would expect if the different facets grew at random perpendicular velocities. These observations suggest that there is some deterministic physical mechanism that guides the formation of trigonal crystals, and this paper is mainly focused on understanding this mechanism.

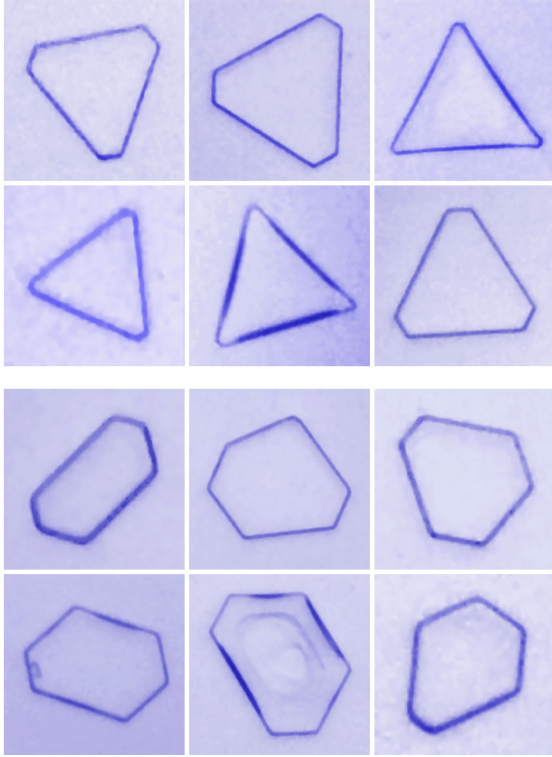

Figure 5: Photographic examples of small platelike snow crystals grown at -10 C in a free-fall convection chamber [2009Lib4]. Basic hexagonal crystals were the norm in this data set, and trigonal forms (top six examples) were the most common low-$H_1$ crystals, as seen in Figure 4. Irregular forms (bottom six examples) were relatively rare.

## Trigonal Snow Crystals in the Laboratory

In perhaps the first laboratory study focusing on trigonal snow crystals, Yamashita [1973Yam] surveyed earlier observations and examined the probability of finding trigonal crystals at different temperatures in a free-fall cloud chamber, where crystals grew in air with a background supersaturation roughly equal to that of supercooled water droplets. These data revealed sharp peaks in the formation of trigonal crystals near -7 C and -25 C, with up to 60 percent trigonal forms at both peaks. These fractions are higher than observed by subsequent researchers, but this may depend on the $H_1$ cutoff used to separate trigonal crystals from hexagonal forms. Irregular forms were less common than trigonal crystals in this study, supporting earlier photographic observations of atmospheric snow crystals. Yamashita also pointed out that many larger snow crystals likely went through an early trigonal phase before growing out to nearly hexagonal morphologies, as indicated by trigonal surface markings near the crystal centers (see also Figure 3).

Takahashi et al. [1991Tak] produced another substantial data set of snow crystals growing in air with a supersaturation near that of liquid water, reporting that trigonal crystals were especially prevalent near -4 C and -8 C, but with lower probabilities than [1973Yam]. Libbrecht et al. [2008Lib1] again found relatively low numbers of trigonal crystals over a broad temperature range, reporting that simple trigonal plates were especially prevalent with low supersaturations at -2 C and -10 C [2009Lib4].

The various observations, both in nature and in the lab, convincingly show that trigonal snow crystals can be found over a broad range of growth conditions. The statistical analysis in [2009Lib4] confirms that trigonal forms are more common than one would expect if the six prism facets all grew at different but randomly determined rates. These data all describe growth in air at a pressure near one atmosphere, and, to my knowledge, there have been no reports of trigonal forms appearing in low-pressure growth data. Unfortunately, the relative paucity of laboratory data does not provide an especially clear picture of how trigonal growth varies with temperature and perhaps nucleation, and there are little data investigating growth as a function of supersaturation.

What these earlier measurements lack are unambiguous clues as to the underlying physical processes that promote trigonal growth over the usual hexagonal morphologies. Most snow crystal phenomena are largely deterministic, so a careful experiment should, in principle, provide a dependable and reproducible sample of trigonal forms. Indeed, the laboratory data do indicate that trigonal forms appear more readily under certain conditions, although the published observations are not consistent with one another to a satisfactory degree, making it difficult to ascertain just what optimal conditions most reliably result in the creation of trigonal forms.

This situation changed somewhat when I began an experimental investigation into the growth of platelike snow crystals on ice needles at -14 C [2020Lib1]. During a series of observations at this temperature, I discovered that trigonal crystals grow quite readily – with nearly 100 percent probability – over a remarkably narrow range in background supersaturation (denoted $\sigma_\infty$, which is generally much higher and better determined than $\sigma_{surf}$ at the crystal surface). I had not looked with sufficient resolution in supersaturation to notice this localized phenomenon earlier, but it prompted the detailed investigation of trigonal growth on ice needles presented here.

## ❄ Trigonal Snow Crystal Growth on Ice Needles

To begin with some background on the peculiar physics of ice crystal growth from water vapor, I recently developed a Comprehensive Attachment Kinetics (CAK) model that can explain the abrupt changes in snow crystal growth as a function of temperature often summarized in the Nakaya diagram [2021Lib, 2019Lib1]. A key feature of the CAK model is the concept of Structure-Dependent Attachment Kinetics (SDAK), in which the attachment kinetics on faceted surfaces may depend on the mesoscopic structure of the crystal, especially the width of a faceted surface [2003Lib1, 2021Lib]. A microscopic physical mechanism that can produce the SDAK phenomenon is described in detail in [2021Lib, 2019Lib1]. The CAK model has been well supported by subsequent targeted experimental investigations [2019Lib2, 2020Lib, 2020Lib1, 2020Lib2], suggesting that many of the overarching tenets of the model provide a reasonably accurate physical picture of snow crystal growth dynamics over a broad range of environmental conditions.

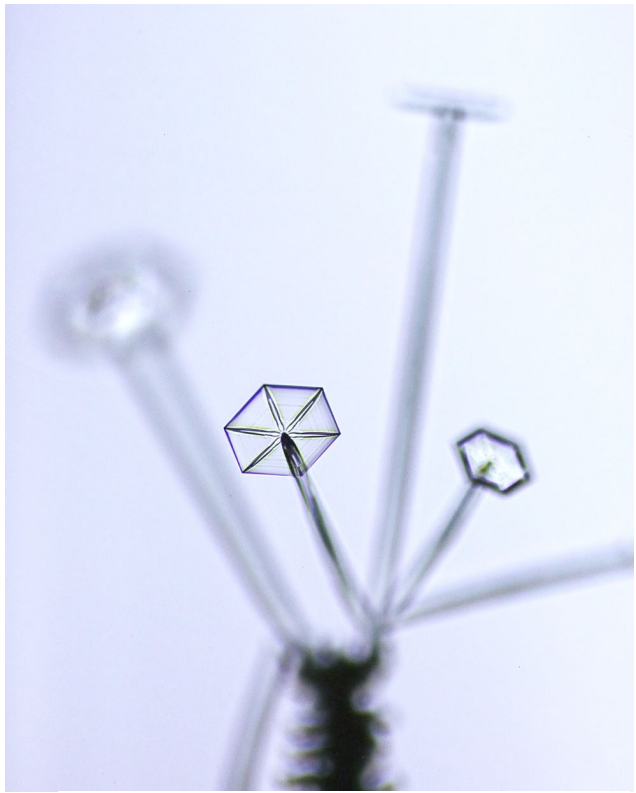

Figure 6: To obtain this photograph, a set of c-axis "electric" ice needles was first created on the end of a frost-covered wire (the black structure at the bottom of the frame). The slender needles were then transported to an attached chamber where hexagonal platelike snow crystals grew on the needle tips. Here the plate diameters are roughly 200 μm. The wire can be rotated about the vertical axis to bring each plate into focus for closer observation.

To further test the CAK model, I have been looking with particular care at snow crystal growth on slender ice needles at -14 C, because the SDAK effect on prism facets is especially strong at this temperature, leading to an Edge-Sharpening Instability (ESI) that promotes the formation of thin plates [2021Lib, 2020Lib1]. In a nutshell, the prism attachment coefficient $\alpha_{prism}$ depends especially strongly on facet width at -14 C, because the effective nucleation barrier drops precipitously as the facet width decreases to molecular scales at that temperature. The CAK model provides a physical foundation for understanding the dynamics of snow crystal growth as well as a narrative and nomenclature for better discussing both the formation of thin plates and trigonal platelike crystals on ice needles.

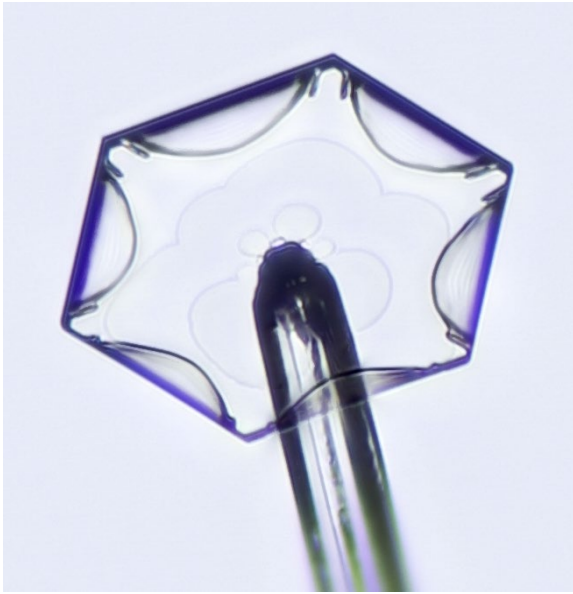

Figure 7: This image shows closer view of a hexagonal platelike snow crystal growing on the end of an ice needle. The plate diameter is about 400 μm. The top basal surface is essentially flat and featureless, presumably with a slightly concave structure (not visible in the photo) required for diffusion-limited faceted growth. The hexagonal tips show the initial development of "ridge" structures [2021Lib], which form on the underside of the plate. (Figure 6 shows narrower ridges that grow at somewhat higher supersaturations.) The faint "billowing" features near the plate center are macrosteps that form where the lower basal surface meets the needle tip, as there is no basal nucleation barrier at that location. These macrosteps propagate outward as the crystal grows, which is easily seen in timelapse videos. The c-axis needle is initially quite slender, but then thickens with time with an overall hexagonal columnar structure.

The apparatus used to make these observations is a dual-diffusion-chamber system described in [2014Lib1, 2021Lib]. The first diffusion chamber was used to produce "electric" c-axis ice needles by applying a high voltage to a frost covered wire, with the wire tip placed at -6 C in highly supersaturated air. Once the ice needles grew to 1-2 mm in length, the voltage was removed and the needles were transported to an adjoining diffusion chamber to observe their growth under adjustable and carefully controlled growth conditions, always in air at one atmosphere. Figure 6 shows a typical set of ice needles, here with thin hexagonal plates growing on the ends of the needles. Figure 7 shows a closer view of a single plate from a different set of needles.

Using slender ice needles as seed crystals has a distinct experimental advantage compared to normal seed crystals, in that there is only a single basal surface initially present at the needle tip, accompanied by a single set of basal/prism corners [2021Lib]. This provides a simpler platform for investigating the formation of thin platelike crystals, as the diffusion-limited growth dynamics on a typical small seed crystal is significantly complicated by the presence of two basal surfaces with competing sets of basal/prism corners.

Figure 8 shows the results of a single data run in which the supersaturation was slowly lowered as a function of time over several hours while new needle crystals were periodically grown and inserted into the second diffusion chamber for observation. The resulting platelike crystals clustered mainly near the trigonal line, so the value of $H_1$ is roughly equal to $H_1 \approx 1 - T_1$ for each data point. Note that this is a complete data set for this run, where every crystal with a reasonably well-formed morphology was recorded. (A few asymmetrical, clearly abnormal structures were not counted.) Additional data runs showed that the general trends seen in Figure 8 were quite reproducible.

At high $\sigma_\infty$ in Figure 8, thin hexagonal plates grew on the ends of all the ice needles, and Figures 6 and 7 show examples of this happening. As $\sigma_\infty$ dropped below about 8 percent, however, trigonal plates appeared instead, including some nearly perfect triangular plates with $T_1 \approx 1$. Remarkably, essentially no hexagonal plates (depending on where one places the $T_1$ cutoff) grew along with the trigonal plates. When $\sigma_\infty$ was reduced below 6 percent, platelike crystals no longer grew from the ice needles, and the initially slender ice needles grew slowly into blocky forms or into simple columns. Figure 9 shows several examples of platelike crystal morphologies with different values of $T_1$.

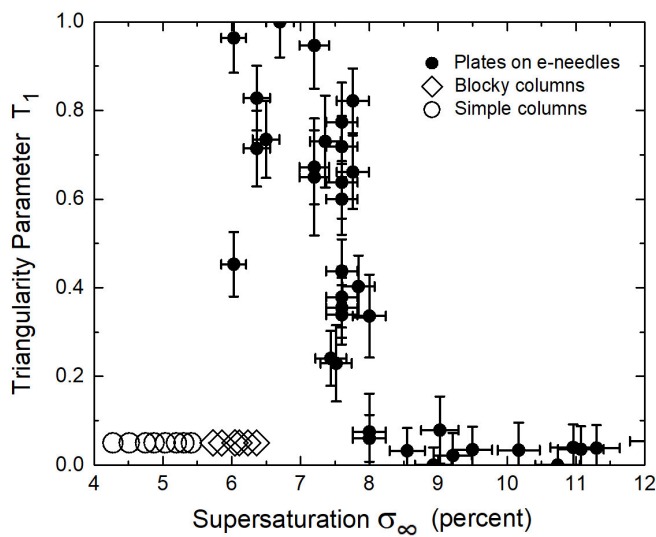

Figure 8: A data run in which the supersaturation $\sigma_\infty$ was slowly lowered over several hours while crystal growth on needle crystals was observed at -14 C. To avoid selection bias, all growing crystals were recorded during this run (except for a few malformed outliers). The data clearly show a transition from hexagonal plates to trigonal plates as $\sigma_\infty$ decreases, followed by additional transitions to blocky forms and then to simple columns.

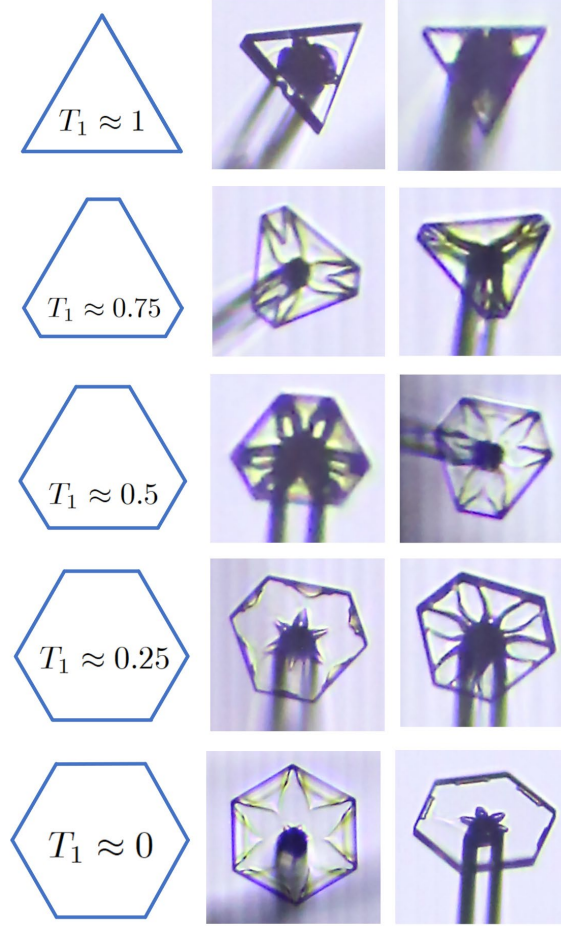

Figure 9: Examples of platelike crystals growing on ice needles at different measured values of $T_1$. Note that the crystals are not viewed face-on, as evidenced by the orientation of the c-axis ice needles.

The behavior seen in Figure 8 provides an important clue for understanding the formation of trigonal plates at -14 C. The "SDAK dip" on the prism facet is most pronounced at -14 C [2020Lib1], meaning that the Edge-Sharpening Instability (ESI) responsible for plates emerging on the end of ice needles is especially sensitive to growth conditions. I observed trigonal plates forming at other temperatures near -14 C, but the drive to form trigonal plates appeared to be strongest at precisely the same temperature that plate formation is also strongly driven. This is almost certainly not a coincidence, and it means that the ESI responsible for plate growth near -14 C likely plays an important role in the development of trigonal forms. This correspondence does not explain every aspect of trigonal growth, specifically at other temperatures, but it provides an important clue for ascertaining the physical mechanisms underlying trigonal growth from ice needles at this specific temperature.

## Triangular Plates

Figure 10 shows the development of a nearly perfect triangular snow crystal on an ice needle. In this example, the triangular structure emerged quite quickly when the crystal was too small to image clearly, and the equilateral-triangle morphology was maintained as the crystal grew larger. This series of images provides another important clue for understanding triangular growth, because all the triangular crystals I observed seemed to develop very quickly, when the plate was quite small. Once a trigonal crystal was observed with $T_1 < 1$, it never developed into a triangular crystal with $T_1 \approx 1$. This was true not only for full triangular crystals, but also for individual triangular corners (that is, a sharp corner consisting of a single, exceedingly small prism facet flanked by two larger prism facets).

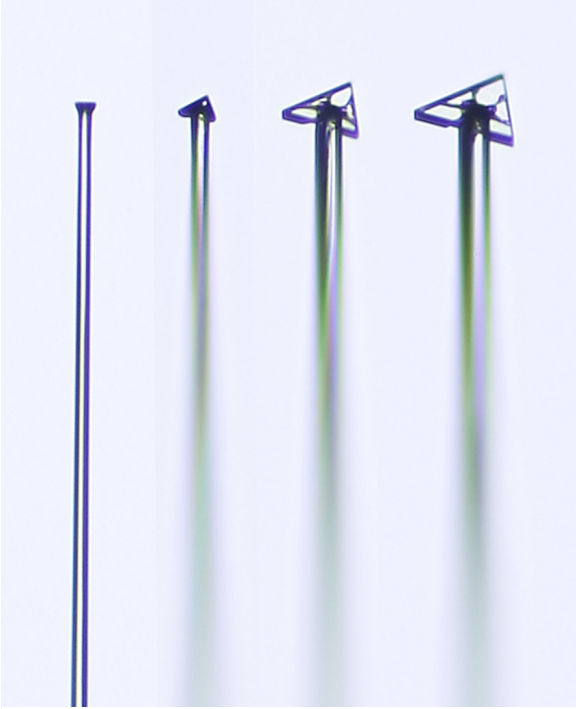

Figure 10: This series of photos illustrates the development of a nearly perfect triangular plate on the end of an ice needle. The first image shows a view perpendicular to the c-axis needle, showing the plate edge-on. The other images show the same crystal from a different angle to better view the plate morphology as it grows. In all my observations, I never witnessed a trigonal plate with $T_1 < 1$ transforming into a triangular plate with $T_1 \approx 1$. Rather, the three sharp points on triangular crystals always emerged early, when the overall plate size was 20 µm or less. Conversely, triangular plates frequently transformed into trigonal plates with $T_1 < 1$, which occurred both when $\sigma_\infty$ was above or below the optimal value that maintained triangular growth.

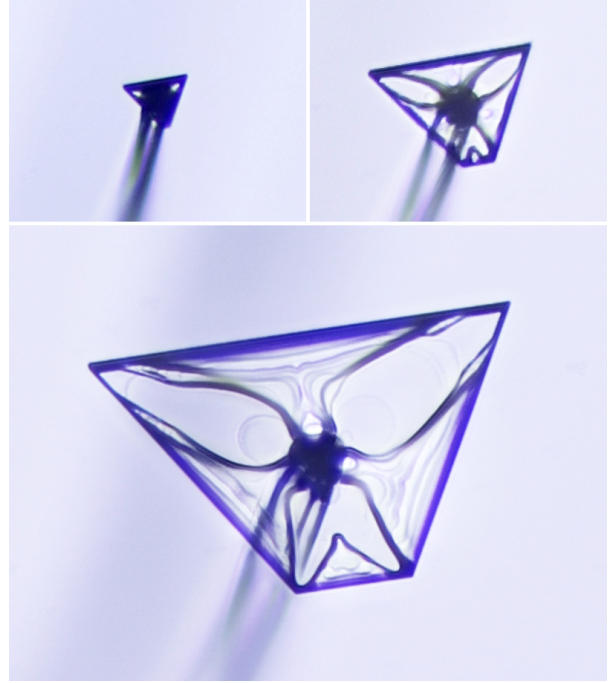

Figure 11: In this example, a plate with two sharp triangular corners develops over time. The longest prism facet in the bottom image is 320 µm in length, and the top images are shown at the same scale. The two main triangular tips developed early and maintained their sharp morphology as the crystal grew, over an elapsed time of 33 minutes in this series. The bottom facet, however, did not develop into a third sharp triangular tip.

Figure 11 shows an example where two sharp triangular tips appeared when the crystal was small and maintained their sharp-tip morphology as the crystal grew. The third tip did not establish itself fully when the crystal was small, and it retained its truncated-tip morphology during further growth. Both Figures 10 and 11 show that sharp triangular tips (where a prism facet with near-zero width is flanked by two large prism facets) can grow stably under the right conditions once this morphology is established. With plate-on-needle growth at -14 C, however, the sharp-tip morphology seems to develop only when the plate first emerges from the columnar needle, and not at later times.

# Diffusion-Limited Growth and Anisotropic Attachment Kinetics

The growth of a sharp-tipped triangular snow crystal necessitates some degree of anisotropic attachment kinetics to maintain, specifically where $\alpha_{prism}$ on the tip facet must be higher than $\alpha_{prism}$ on the much larger neighboring facets. Diffusion-limited growth in the absence of this kind of anisotropic attachment kinetics will not yield a stably growing tip structure. To see this, consider the region near a sharp tip illustrated in Figure 12.

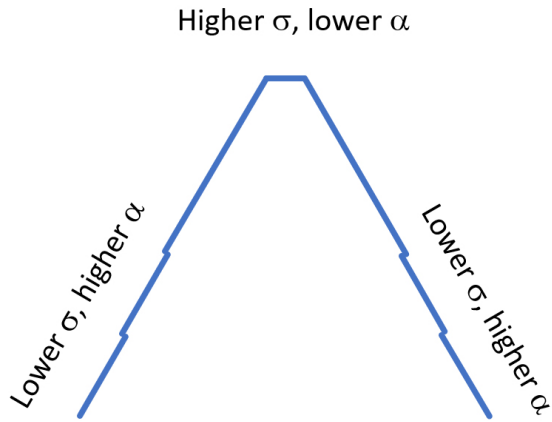

Figure 12: A sketch of a growing triangular crystal near one of the sharp tips. Far from the tip, terrace steps provide a higher attachment coefficient, thus balancing the lower supersaturation to yield normal faceted growth. Near the tip, however, the top prism terraces will be step-free, so some other form of anisotropic attachment kinetics on the different prism facets is necessary to maintain the sharp-tip structure.

Considering the large prism facets first, these maintain an essentially flat structure as they grow, requiring a constant perpendicular growth velocity at all points on the facet. Particle diffusion produces a larger $\sigma_{surf}$ at the tip compared to the large-facet centers, however, so faceted growth requires that $\alpha_{prism}$ be larger at the facet centers. The $\alpha_{prism}$ needed for stable constant-velocity growth is provided by molecular steps on the large-facet surface, and the step density automatically adjusts itself to maintain a perpendicular growth velocity that is equal at all points on the facet surface. The large facets are thus always somewhat concave in shape, although these surfaces appear flat in photographs. This self-regulating process is well known as the primary mechanism for maintaining a faceted morphology when the growth is limited by diffusion.

The situation becomes more interesting near the sharp tip, however, where all three prism facets exhibit their top molecular terraces (see Figure 12). Nucleation on these top terraces, free from molecular steps, is generally what limits the growth of the entire faceted structure. In the case of a sharp tip structure, the top terraces on all three facets will be positioned quite close to the tip. Because the length scale over which $\sigma_{surf}$ changes appreciably will likely be much larger than the spacing between the three top terraces, one can assume that $\sigma_{surf}$ is essentially constant near the growing tip.

From simple geometrical considerations with a 60-degree tip, the narrow prism facet at the apex in Figure 12 must grow at exactly twice the perpendicular growth velocity of the adjacent large facets. Moreover, because $\sigma_{surf}$ is nearly constant at the tip, this means that $\alpha_{prism,tip}$ on the top terrace of the apex facet must be about double that on the top terraces of nearby large facets. In fact, the stable tip growth suggests $\alpha_{prism,tip} > 2\alpha_{prism,large}$, as this allows the width of the top terrace to automatically adjusts itself until the proper 2:1 growth velocity is achieved via the Gibbs-Thomson effect. Explaining a factor-of-two difference in the attachment coefficients on these surfaces is a nontrivial challenge, however, because the molecular structures of all three of these top prism terraces are identical. This is the crux of the trigonal-growth problem – explaining how ostensibly identical prism surfaces can have markedly different attachment kinetics.

I believe that Structure-Dependent Attachment Kinetics (SDAK) can again provide a ready explanation for the anisotropic attachment kinetics that is required to maintain a growing sharp-tip structure. In the SDAK model I have proposed [2003Lib1, 2019Lib1, 2020Lib1, 2021Lib], $\alpha_{prism}$ depends on the width of a growing prism facet at -14 C, because the broad-facet nucleation barrier is greatly diminished by surface-diffusion effects. This SDAK model is essentially a two-dimensional (2D) effect, operating to promote the growth of thin platelike crystals with sharp prism edges.

For triangular snow crystals at -14 C, I envision a three-dimensional (3D) version of the same SDAK effect, this time operating on prism facets that are small in both lateral dimensions (as opposed to a plate edge, where the prism facet is only small in one dimension). Given that the 2D SDAK effect yields large increases in $\alpha_{prism}$ for prism edges at -14 C [2020Lib1], it would not be surprising to see an increase of a factor of two or more in $\alpha_{prism}$ on the sharp triangular tip, where the top prism terrace becomes a small island. Developing a comprehensive model of this new 3D SDAK effect is a nontrivial task but hypothesizing its existence to explain the growth of triangular plates at -14 C seems like a reasonable next step toward a fuller understanding of the complex phenomenon of snow crystal growth.

Continuing this reasoning, I hypothesize that the 3D SDAK effect then yields a Tip Sharpening Instability (TSI) that corresponds to the Edge Sharpening Instability (ESI) resulting from the 2D SDAK effect. As with my previous efforts to understand the CAK model including the SDAK effect, creating an unambiguous molecular model of all the underlying physical processes will likely remain an unsolved problem for many decades. There is hope that molecular-dynamics simulations will shed some light on these issues, and we have seen recent progress on this front [2021Lib]. In the meantime, however, I have found it quite beneficial to develop the semi-empirical CAK model using qualitative physical insights coupled with targeted experimental investigations. At the very least, this avenue of model development suggests additional targeted experiments that can shed light on the underlying model assumptions. I proceed with the SDAK model hypothesis, therefore, as it provides a framework for guiding our discussion of trigonal snow crystals.

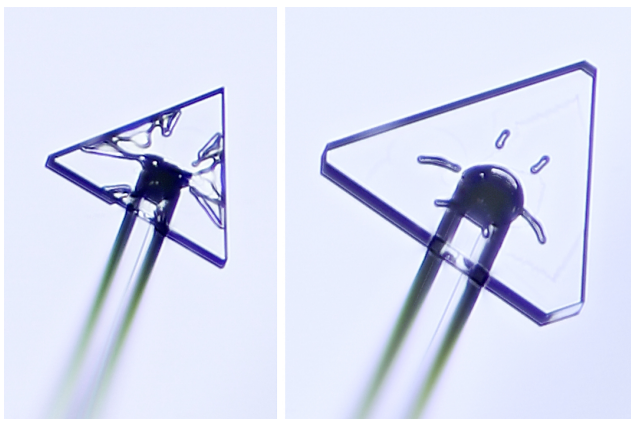

Figure 13: Lowering the supersaturation sufficiently will disable the Tip-Sharpening Instability so sharp triangular tips develop into larger prism facets. The image on the left shows a near-perfect triangular plate with prism facet lengths of 165 μm. The surface markings on the underside of the plate appear to have a negligible effect on its overall triangular structure. The image on the right shows the same crystal (at the same scale) after an additional 35 minutes of growth after lowering the supersaturation. The resulting slower growth deactivates the TSI mechanism, so then all six prism facets grow at roughly equal rates, causing $T_1$ to decrease with time.

## Morphological Development of Triangular Plates

While I have never witnessed a non-sharp tip on an established platelike trigonal crystal develop into a sharp triangular tip, I have frequently observed sharp triangular tips evolve into trigonal shapes where the tips are not sharp. Lowering the supersaturation will easily produce this evolution, and Figure 13 shows one example of this happening.

As with the ESI, the TSI will only operate above some threshold supersaturation, below which $\alpha_{prism}$ on the sharp-tip facets will be less than twice that of $\alpha_{prism}$ on the larger facets. When this happens, the plate morphology will slowly evolve to lower $T_1$ values. If $\alpha_{prism}$ becomes equal on all six facets, then the shape will slowly evolve toward that of a simple hexagon, even if it started out with a trigonal shape.

This overall behavior reflects the general maxim in diffusion-limited snow-crystal growth that anisotropic growth morphologies on large scales require anisotropic attachment kinetics [2021Lib]. As discussed with Figure 12, diffusion alone, specifically the Berg effect, will not promote the growth of triangular plates, but will instead result in evolution toward a basic hexagonal shape. The observation demonstrated in Figure 13, therefore, supports the idea of a new form of SDAK at the triangular tips, resulting in a Tip-Sharpening Instability that will drive the formation of triangular plates.

Increasing the supersaturation will also cause triangular plates with $T_1 \approx 1$, to evolve into trigonal plates with $T_1 < 1$, and this phenomenon is illustrated in Figure 14. My interpretation of these images is that a TSI was initially present, yielding the two sharp triangular tips seen in the first image. As described above, the TSI was necessary to maintain the sharp-tip structure during small changes in $\sigma_\infty$ with time. Increasing $\sigma_\infty$ substantially then caused an abrupt jump in the growth rates of the broad facets, while having a smaller effect on the tip growth. This changed the growth ratio sufficiently that the tip was no longer advancing at twice the rate of the broad facets. The narrow tip then broadened so the TSI was inoperable and $T_1$ soon dropped below unity.

A better analysis would require full 3D numerical modeling of this complex structure, which is not yet possible with existing techniques. Nevertheless, the example does illustrate that increasing the supersaturation

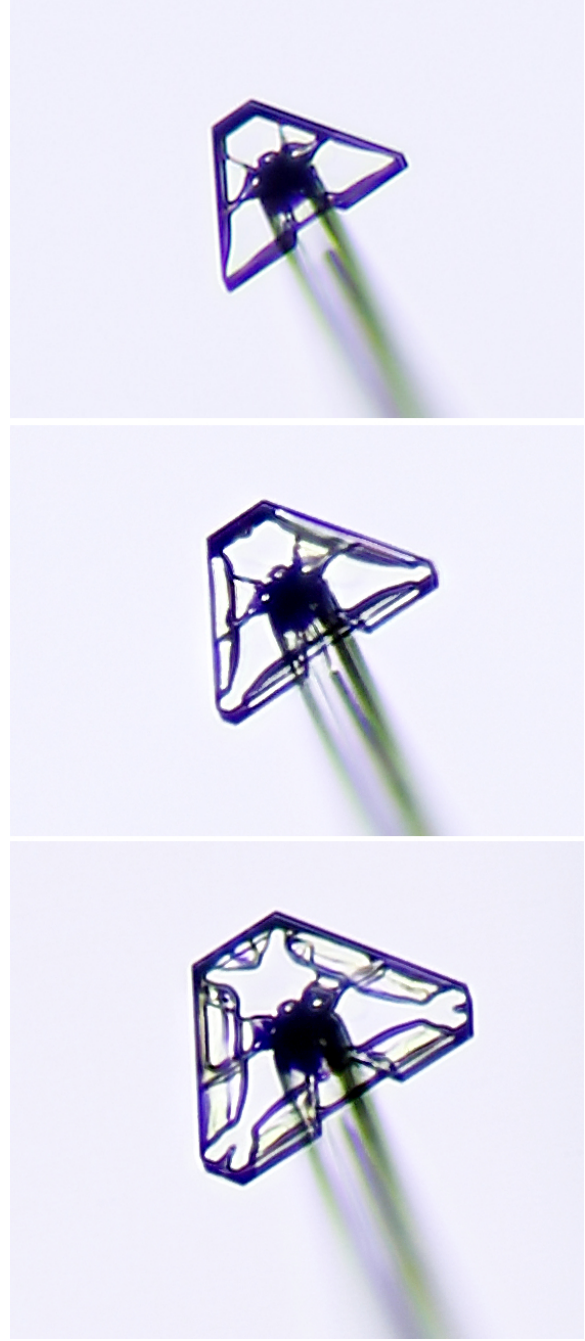

Figure 14: This series shows how triangular tips can evolve to larger facets when the supersaturation is increased. The top image shows a partial triangular plate (with two sharp tips) growing at $\sigma_\infty \approx 7\%$. Turning the supersaturation up to $\sigma_\infty \approx 12\%$ yielded the second and third images after additional growth, showing ridges developing at the sharp tips and thin splitting as the sharp tips evolved into larger prism facets.

can disrupt the steady-state growth of a sharp-tipped triangular crystal.

From these and other observations not presented here, it appears that triangular plates forming on c-axis needles at -14 C can be understood as arising from a 3D form of Structure-Dependent Attachment Kinetics (SDAK), as this could provide a sufficiently high degree of anisotropy in the attachment kinetics. Specifically, the growth of triangular crystals with $T_1 \approx 1$ requires that $\alpha_{prism}$ on the sharp tips be at least twice that of $\alpha_{prism}$ on the large facets. The required SDAK effect then leads to a Tip Sharpening Instability that establishes and maintains the sharp triangular tips. This physical effect is a somewhat natural extension of the Comprehensive Attachment Kinetics model [2021Lib, 2019Lib1], thus further supporting that model and its explanation of the Nakaya diagram.

## Morphological Development of Trigonal Plates

Trigonal plates span a broader spectrum of possible morphologies, covering the whole trigonal line in the $(H_1, T_1)$ plane, so any analysis will not be as clean as with the extreme case of sharp tips on triangular crystals. Moreover, many trigonal morphologies depend on the entire growth history of the crystal, which further complicates the discussion. Nevertheless, Figure 15 shows one illustrative example of a basic trigonal plate and how its growth changed when the supersaturation was increased.

In the first image in Figure 15, the trigonal form grows in an essentially stable fashion with the short facets having a 60% higher perpendicular growth velocity than the long facets. One can perform a qualitative diffusion analysis like that in Figure 12, except this time focusing on a small region around a single corner between a short and long facet. Once again, because the top facet terraces are both quite close to the corner, one expects that the surface supersaturation will be nearly identical on each. The difference in growth velocities, therefore, indicates that $\alpha_{prism}$ on the short facet must be about 60% higher than that on the long facet. And the TSI no longer applies in this case because both top prism terraces are long and narrow.

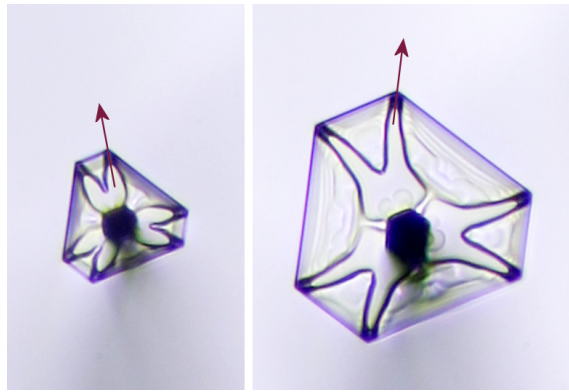

Figure 15: The image on the left shows a trigonal plate with a relatively high $T_1$, and measurements reveal that the short facets have a perpendicular growth velocity about 60% faster than the long facets. The growth ratio can also be ascertained from the ridge structure, as each ridge traces the position of a corner between facets as a function of time. After this first picture was taken, the supersaturation was increased and the photo on the right was taken after additional growth. Now the long and short facets have similar overall appearances, and the ridges indicate that they have nearly equal growth rates.

One possible explanation for the difference in $\alpha_{prism}$ is that the short facet is slightly thinner than the long facet, so we can then invoke the normal 2D SDAK effect. The two sets of facets do have a slightly different appearance in Figure 15, but this is not sufficient to say much about the top terrace widths. Nevertheless, the measurements in [2020Lib1] show that $\alpha_{prism}$ at -14 C increases roughly 100-fold as thin edges of platelike crystals develop, so it would not be surprising to see a 60% increase in $\alpha_{prism}$ between the short and long facets in Figure 15.

Raising the supersaturation (second photo in Figure 15) increases the growth rates of both the long and short facets, but the long facets started out slower and exhibit a larger net velocity increase compared to the short facets.

As with triangular crystal in Figure 14, it appears that the higher $\sigma_\infty$ reduced the $\alpha_{prism}$ difference between facets, equalizing the growth rates and lowering $T_1$ in the process. Soon, as seen in Figure 15, the long and short facets develop similar appearances and grow at similar rates. Extending this growth behavior forward, the overall shape would evolve to become more hexagonal with time.

My main conclusion from this example is that a substantial series of quite careful measurements would be required to draw any quantitative conclusions along these lines. Moreover, the ridge structures present on these platelike crystals may substantially affect the growth dynamics and edge thicknesses. Understanding this better would thus require a full 3D numerical analysis of the growing crystals, which is not technically possible at present.

A secondary conclusion, however, is that the trigonal growth behavior seen in this example is at least consistent with the basic tenets of the proposed SDAK mechanism. The crystal morphology may be too complex for detailed analysis, but the overarching behaviors seen in this and other trigonal plates-on-needles generally supports the CAK model interpretation.

## ❄ A Physical Model Explaining the Growth of Trigonal Snow Crystals on Ice Needles at -14 C

A fundamental mystery surrounding trigonal snow crystals since their first observations relates to how they form and develop. Focusing on platelike crystals, the basic hexagonal morphology is certainly expected because the six prism facets are all essentially identical and would be expected to exhibit identical growth velocities. And, indeed, simple hexagonal forms are commonly found whenever platelike crystals grow.

But why would a trigonal plate form, especially a sharp-tipped triangular plate? What physical processes differentiate the growth rates of the short and long prism facets, and why do they alternate around the crystal to yield a trigonal shape? With these new observations of trigonal plates growing on ice needles, I have created a developmental model that can explain the many of the overarching features of how these enigmatic crystals originate and develop.

There have been some previous attempts to explain the formation of trigonal snow crystals, but I feel that they fall short of achieving a reasonable understanding of the underlying physical processes. For example, Murray et al. [2015Mur] recently suggested that trigonal ice crystals in the atmosphere may not be made from normal crystalline ice Ih, but rather from a stacking-disordered ice lattice with trigonal symmetry. This seems quite unlikely, in my opinion, especially when cubic ice Ic – the only other form of ice clearly identified under atmospheric conditions – has never been observed in single-crystal form (to my knowledge). Moreover, small snow crystals are often essentially dislocation-free single crystals, and even twinning is relatively rare. It is difficult to imagine the occurrence of near-perfect stacking disorder in macroscopic snow crystals. X-ray crystallographic measurements could prove the existence of a triangular lattice structure, and potential trigonal samples for such a study are not hard to fine. Until such evidence appears, however, I consider the stacking-disorder hypothesis somewhat improbable.

Libbrecht and Arnold suggested that aerodynamic stability considerations could result in trigonal forms [2009Lib4], but I now believe that this is not likely a viable explanation either. In a subsequent investigation of the aerodynamics of snow crystal growth more broadly, I found that diffusion effects on snow crystal growth (also known as the ventilation effect [1982Kel]) are probably too insignificant around small snow crystal plates to affect their early growth dynamics [2009Lib3]. Moreover, the reasoning in Figure 12 suggests that anisotropic attachment kinetics are needed for trigonal growth, which is separate from aerodynamics considerations. Plus, of course, aerodynamic effects are completely absent in the plate-on-needle observations described above, which take place in static air.

In thinking about this problem more carefully, I found that it was not difficult to conjure up physically plausible models that could facilitate the formation of sharp triangular tips. Given that these tips will grow stably via the TSI, assuming the 3D SDAK effect described above, all that is required is some sequence of events that yields three such tips in an alternating pattern to make a triangular crystal. A major problem with most such scenarios, however, is that non-triangular forms should also appear with sharp triangular tips, especially the diamond-shaped form illustrated in Figure 16. Both diamonds and triangles possess stable triangular tips, and both these forms would grow stably with the TSI. However, observations clearly reveal that triangles form much more readily than diamonds. Explaining this disparity proved to be an important consideration in the formation problem. If sharp triangular tips emerged somewhat randomly, one would expect to see diamonds appearing with some regularity, whereas few to none are observed in experiments or in nature. This observation suggests that there must be some deterministic path that guides the formation of triangular plates with much higher probability than diamond-shaped plates.

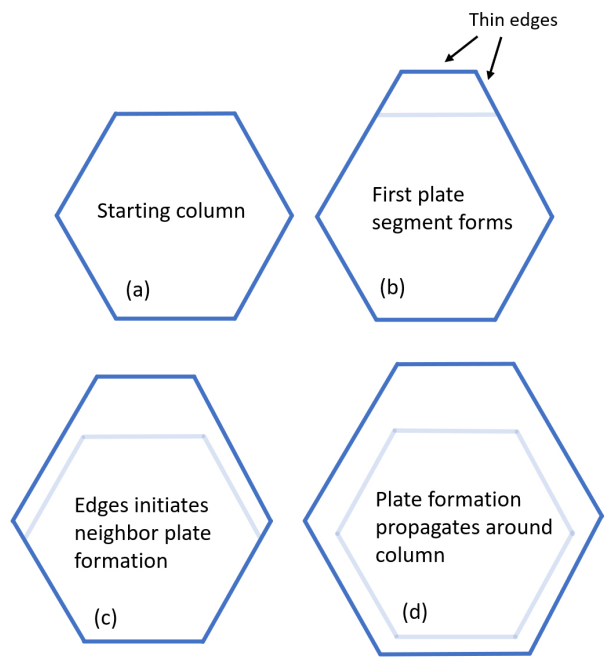

Figure 17: This series of sketches illustrates how a thin hexagonal plate can emerge from the tip of a hexagonal needle crystal. As described in the text, the SDAK mechanism and the resulting ESI naturally cause a plate segment to form at some random location, and the plate segment will quickly "wrap around" the column to form a full hexagonal plate. The slight initial deviation from a perfect hexagonal shape will soon become insignificant as the six facets continue growing outward at nearly identical rates.

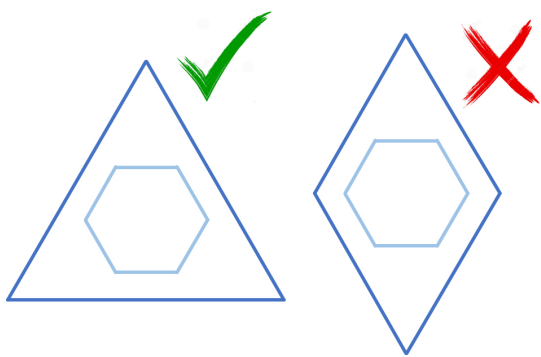

Figure 16: Triangular snow crystals grow readily from hexagonal needle precursors at -14 C (left), but diamond-shaped crystals clearly do not (right). Explaining this disparity is an important and nontrivial requirement for any physical model describing the formation of triangular plates on ice needles.

## Origin and Development of thin Plates on Ice Needles

Before describing a model for the formation of trigonal plates on ice needles, it is beneficial to consider first how hexagonal plates arise, which is illustrated in Figure 17. The first sketch (a) in this figure depicts the initial ice needle, created using high electric fields in our dual-diffusion-chamber apparatus [2014Lib1] and then transported (with the electric fields removed) to the second chamber for observation. The radius of this initial column is no more than a few microns at its tip.

As the needle grows, its radius increases and the basal/prism edges on the needle tip will develop thin plates via the ESI, being driven by the SDAK phenomenon, which is part of the CAK model of snow crystal growth [2021Lib, 2019Lib1]. I assume that the CAK model is correct in this discussion, and note that the growth of such plates, once established, is described in some detail in [2020Lib1].

Referring to Figure 17 (b), we assume that the ESI does not produce a thin plate on all prism edges simultaneously, but instead a single plate segment likely first appears at random on one of the basal/prism edges. Once the plate segment forms, $\alpha_{prism}$ increases dramatically on the thin prism facet, as this is the nature of the underlying SDAK effect. This increases the plate edge growth relative to the broad needle facets, so the plate segment grows out rapidly from the needle, as illustrated in Figure 17 (b).

As the plate segment extends out from the column, we see that three plate edges soon develop – the central edge flanked by two side edges. Because the side edges are also quite thin, $\alpha_{prism}$ will soon become high on these also, via the SDAK effect, so the side edges will begin to grow outward, causing the plate segment to "wrap around" the column, as shown in Figure 17 (c). Put another way, the narrow side edges have a much-reduced nucleation barrier compared to the broad needle facets, and this enhanced nucleation yields a train of molecular steps that propagate onto the needle facets.

This process continues with the other prism surfaces, so the plate quickly wraps around the column completely, yielding a full platelike crystal on the end of the columnar needle. Once the six segments start growing outward at roughly equal rates, the plate evolves to become essentially hexagonal in shape, so it is no longer obvious that one segment had a brief head-start relative to the others. The result is a near-perfect hexagonal plate growing out from the ice needle, which is commonly observed.

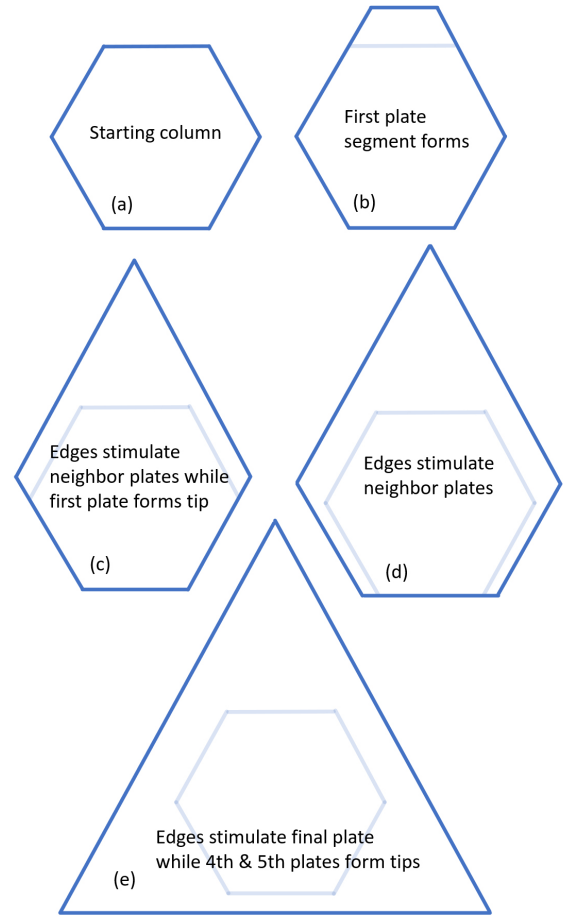

Figure 18: This series of sketches illustrates how a thin triangular plate can emerge from the tip of a hexagonal needle crystal. The main difference from Figure 17, as described in the text, is that the wrap-around process proceeds more slowly when the supersaturation is so low that the ESI is just barely operational. In this case, plate segments can grow out to become triangular tips before the neighboring plate segments develop. If the timing is right, this process can yield full triangular plates, as are observed. If the timing is somewhat off, "failed" triangular plates with only one or two triangular tips can result, and these are also commonly seen in observations. It would be highly unlikely for a diamond-shaped crystal to result from this process, however, thus explaining why diamond-like shapes are not seen in plate-on-needle data.

Next consider the formation of a plate when the supersaturation is somewhat lower, so the ESI is weaker and just barely able to produce the transition from a basal/prism edge to a thin plate. The formation of the first plate segment eventually takes place, and this is shown in Figure 18 as the transition from (a) to (b). The faster growing plate will tend to lower the supersaturation around the crystal, however, so the first "wrap-around" event may be quite slow. While this is happening, the first plate segment may grow out so far that it becomes a triangular tip, activating the TSI because of its small apex width, as illustrated in Figure 18 (c). Note that, because the plate segment is attached to the needle, the plate thickness will be smallest on the outermost facet, which facilitates the tip formation. Once the tip forms, the TSI then ensures that it will grow stably outward thereafter.

After some additional growth, the nascent plate will continue to wrap around the perimeter of the needle, proceeding from (c) to (d) in Figure 18. Once again, however, it will take time to make these transitions with the weakened ESI, and the final step to the sixth facet may occur after the 4$^{th}$ and 5$^{th}$ plate segments have grown out to form triangular tips, as indicated in Figure 18 (e). If this chain of events proceeds as indicated, the result will be a full triangular snow crystal plate growing out from the ice needle. From there, as illustrated in Figure 11, the tip morphology can remain stable for long periods, yielding a large triangular plate with sharp tips.

Of course, the process depicted in Figure 18 is likely somewhat fragile, leaving numerous avenues for creating a "failed" triangular crystal with fewer than three sharp tips. Indeed, Figures 11 and 14 show plates with only two sharp tips, which would be a likely failure mode as the plate wraps around the needle. In some circumstances, I have found that these incomplete triangles are more likely than full triangular crystals, which is not surprising with this model.

Importantly, it would be somewhat difficult to create a diamond-shaped plate using the progression shown in Figure 18. A diamond would require a substantial imbalance in the wrap-around steps, or perhaps the simultaneous emergence of plate segments on opposite sides of the initial needle. These scenarios seem somewhat unlikely in this model, thus explaining why diamond-shaped plates are much less likely than triangular plates. Overall, therefore, this model does a reasonably good job explaining, at least qualitatively, why triangular plates and related "failed" triangular plates would be relatively common while diamond-shaped plates would be unlikely.

This model also provides a ready explanation for the sharp peak in the production of triangular crystals as a function of supersaturation at -14 C. A high supersaturation would yield a strong ESI and a fast wrap-around process, thus yielding hexagonal plates with high probability, as is observed above below $\sigma_\infty = 8\%$ in Figure 8. In contrast, a low supersaturation would mean that the ESI is too weak to produce any platelike growth, and this is observed at supersaturations below $\sigma_\infty = 6\%$. Only when the ESI is just beginning to turn on will the conditions be conducive to reliably producing trigonal and full triangular plates, which happens over a narrow range in supersaturations.

A particularly pleasing feature of this trigonal origin scenario is how well it fits with the CAK model. If one embraces the CAK model from the outset and then asks how then plates will emerge from ice needle precursors, the growth behaviors in Figures 17 and 18 emerge quite naturally. The formation of trigonal plates is indeed a deterministic process that becomes likely when the ESI is just turning on, so over a narrow range in supersaturation, as is observed. Thus, although the CAK model was not developed with trigonal growth in mind, it seems to do a satisfactory job explaining many observed behaviors observed in the formation of trigonal plates on ice needles at -14 C.

# ❉ Progress Toward Understanding the Formation of Trigonal Snow Crystals

Since they were first documented over 200 years ago, trigonal snow crystals have remained something of an enduring scientific puzzle. Basic crystallography tells us that the six prism facets on a trigonal crystal have identical lattice structures, yet the alternating long and short facets grow at different rates in trigonal crystals. Equilateral-triangular forms are especially enigmatic in this regard, being the most extreme examples of trigonal snow crystals.

In this paper, I describe a focused investigation of trigonal snow crystals that makes substantial progress toward a better understanding of their origin and growth dynamics. Primary results include:

- I developed a novel "recipe" for growing trigonal and triangular snow crystals on the ends of thin ice needles. Trigonal plates form most readily near -14 C in a narrow range of supersaturations, as shown in Figure 8. The trigonal yield peaks at nearly 100 percent, and sharp-tipped triangular plates can be made quite reliably. This discovery enabled a detailed experimental investigation of the growth behaviors of these crystals under controlled environmental conditions.
- The formation of plate-on-needle trigonal crystals at -14 C in air occurs precisely when the supersaturation is just high enough to drive the formation of thin plates. Blocky crystals appear at lower supersaturations, while thin hexagonal plates appear at higher supersaturations, as observed in Figure 8.
- Triangular plates always appear quickly, as the plate first forms on the needle tip. I have never observed a triangular plate with $T_1 \approx 1$ evolving from an established trigonal plate with $T_1 < 1$. The opposite evolution, from triangular to trigonal, it commonly observed by either raising or lowering the applied supersaturation.
- Basic modeling considerations reveal that diffusion-limited growth alone cannot explain the growth of sharp triangular tips, as illustrated in Figure 12. The three prism facets (defined by their top molecular terraces) exist in close proximity near the tip, so the surface supersaturation must be nearly identical on all three. Yet the tip facet grows at twice the rate as is broad-facet neighbors, thus requiring $\alpha_{prism,tip} > 2\alpha_{prism,large}$. This result supports a general maxim in snow-crystal growth that large-scale morphological anisotropy requires a corresponding anisotropy in the attachment kinetics [2021Lib].
- The Comprehensive Attachment Kinetics (CAK) model can provide a physically reasonable explanation for the stable growth of triangular tips. While the 2D SDAK effect describes how $\alpha_{prism}$ on a thin plate edge can be higher than $\alpha_{prism}$ on a broad prism facet [2019Lib1, 2021Lib], a straightforward 3D extension of the SDAK effect could yield the required higher $\alpha_{prism,tip}$ on sharp triangular tips. The 2D Edge-Sharpening Instability (ESI) then turns to a 3D Tip-Sharpening Instability (TSI) on the narrow prism facet, stabilizing the sharp-tipped structure during growth. Creating a full, quantitative 3D model of this phenomenon is a challenging task, but the overarching physical processes in the CAK model provide a plausible physical explanation for the observed stable tip growth.
- I further describe a scenario where triangular crystals can emerge from the tip of a needle crystal via a deterministic process, as illustrated in Figure 18. While the chain of events in this model is speculative, it provides at least one plausible mechanism by which triangular and trigonal forms can emerge in a predictable fashion. Moreover, the model further suggests that diamond-shaped crystals would be rare in comparison to triangular plates, as is observed.

While these results provide a reasonably self-consistent and physically plausible explanation for the origin and growth of triangular and trigonal crystals on ice needles, many questions remain. The investigations presented here all focused on trigonal plates growing on ice needles at -14 C, and the ideas presented in this paper are somewhat limited to this experimental circumstance. Trigonal crystals have also been created in controlled laboratory conditions at other temperatures, however, as summarized at the beginning of this paper, and these likely require one or more different explanations. One particularly pertinent example, continuing with our focus on platelike trigonal crystals, is the relatively common growth of trigonal plates at -2 C [2008Lib1].

Snow crystal growth at -2 C is quite different compared to -14 C, and the SDAK mechanism proposed in the present paper does not apply at the higher temperature. The "SDAK dip" on the prism facet is localized near -14 C [2020Lib1], and its effects are completely absent at -2 C. However, additional growth data at -2 C suggests that there is some new physics at play when surface melting becomes more pronounced near the melting point [2020Lib2, 2019Lib2]. To explain the growth data, I have hypothesized the existence of a second SDAK effect on prism facets at high temperatures, a phenomenon that is, unfortunately, only poorly understood at present. I further hypothesize here that this second SDAK mechanism may play a role in producing platelike trigonal crystals at -2 C.

Much additional study is needed to better understand even the basic behaviors of snow crystal growth at such high temperatures, however, let alone trigonal growth. The data do suggest, however, that there is much left to be learned by additional studies of snow crystal growth at temperatures near the melting point, which is a relatively poorly explored region of phase space.

In summary, I have demonstrated an experimental procedure for readily creating platelike trigonal snow crystals on the ends of ice needles at − 14 C. The technique can be used to produce nearly perfect triangular specimens, which are particularly interesting as an extreme special case. I also describe a qualitative physical model that can explain the stable growth of triangular tips, various aspects of trigonal morphological development, and a deterministic chain-of-events that will plausibly yield triangular crystals, specifically without producing diamond-shaped crystals that are not observed. The phenomenon of trigonal snow crystals, therefore, is perhaps somewhat less puzzling than it was before, at least under these restricted experimental conditions.

## ❋ References